\documentclass[12pt,preprint]{aastex}

\shorttitle{AM-1 and Pal~14}
\shortauthors{Dotter, Sarajedini, \& Yang}
\slugcomment{Accepted by the Astronomical Journal}
\newcommand{\rgc}{\mathrm{R_{GC}}}

\begin{document}

\title{Globular Clusters in the Outer Galactic Halo: AM-1 and Palomar~14 
\thanks{Based on observations with the NASA/ESA {\it Hubble Space Telescope},
obtained at the Space Telescope Science Institute, which is operated
by AURA, Inc., under NASA contract NAS 5-26555, under program
GO-6512.}}

\author{Aaron Dotter}
\affil{Department of Physics and Astronomy, Dartmouth College, 6127 Wilder Laboratory,
 Hanover, NH 03755}

\author{Ata Sarajedini and Soung-Chul Yang} 
\affil{Department of Astronomy, University of Florida, 211 Bryant Space Science Center, Gainesville, FL 32611}

\begin{abstract}
AM-1, at $\sim$120 kpc, and Pal~14, at $\sim$70 kpc, are two of the most distant
Galactic globular clusters known. We present {\it Hubble Space Telescope} WFPC2 
photometry of AM-1 and Pal~14 that reveals unprecedented depth and detail in the 
color-magnitude diagrams of these two clusters. Absolute and relative age 
measurements confirm that both are younger than the inner halo globular cluster 
M~3 by 1.5--2 Gyr assuming all three clusters have similar compositions. 
Thus AM-1 and Pal~14 join Pal~3, Pal~4, and Eridanus (studied by Stetson
et al.) as distant Galactic globular clusters with red horizontal branches
and young ages relative to the inner halo. Within the context of the
entire body of research on the ages of second parameter
globular clusters, the observed correlation between age and horizontal branch
morphology suggests that age is the best candidate for the second parameter.
However, this conclusion is tempered by the lack of precise
chemical abundance determinations for a significant fraction of second parameter
globular clusters.
\end{abstract}


\section{Introduction}
The second parameter phenomenon, a long-standing problem in Galactic astronomy, 
recognizes that the horizontal branch (HB) morphology of Galactic globular 
clusters (GCs) cannot be explained by the first parameter, metallicity, alone: 
at least one more factor, a `second parameter,' must be involved.

In a landmark study, \citet{sz78} demonstrated that GCs inside the solar 
circle (which they deemed the inner halo) exhibit a tight correlation 
between [Fe/H] and HB morphology while those that lie outside the solar circle 
(the outer halo) show no such well-defined trend.  This in turn led \citet{sz78}
to theorize that the formation of the Galaxy was a prolonged, complex process.

While the second parameter phenomenon has informed our understanding of galaxy 
formation, the identity of the second parameter remains a topic of some debate. 
A number of candidates have been suggested including age, helium abundance, CNO 
abundances, 
cluster central density, and stellar rotation (see \citet{fp96} for a review).
It is clear that its influence grows with distance from the Galactic center 
\citep{sz78,zi80,ldz94}. To gain a better understanding of the second parameter it is 
logical to carefully examine the most distant GCs.  Only six known GCs inhabit the 
Galaxy at distances greater than 50 kpc \citep{ha96}
\footnote{References to the Harris catalog actually refer to the February 2003 revision
that can be found at http://www.physics.mcmaster.ca/$\sim$harris/mwgc.dat.}
and five of the six are prime 
examples of second parameter GCs, i.e. metal poor GCs with red HBs.

All six of the GCs with $\rgc >$ 50 kpc: Pal~14, NGC~2419, 
Eridanus, Pal~3, Pal~4, and AM-1 (listed in order of increasing $\rgc$) have been 
observed by the Hubble Space Telescope (HST) using the Wide Field Planetary
Camera 2 (WFPC2).  \citet{ha97} presented a HST/WFPC2 color-magnitude diagram 
(CMD) of NGC~2419---the only one with $\rgc >$ 50 kpc and a blue HB---and 
concluded that it is roughly coeval with the old, metal poor cluster M~92.
In contrast, \citet[hereafter S99]{st99} presented HST/WFPC2 CMDs of Pal~3, 
Pal~4, and Eridanus and concluded that these clusters are younger than relatively
nearby clusters of similar metallicities (M~3 and M~5) by $\sim$1.5--2 Gyr assuming 
the inner and outer halo GCs have similar abundances.

The two remaining GCs, AM-1 and Pal~14, are the subject of this paper.
AM-1 was classified as a GC by \citet{am79} though it had 
been observed previously and named ESO 201-10 by \citet{ho75}. \citet{am79} 
estimated the distance to AM-1 at 300 kpc and, although it has since been 
revised downward to $\sim$120 kpc \citep{aa84,or85,ma89,hi06}, it is still the most 
distant Galactic GC known.  Age estimates for AM-1 have proved uncertain to date
because none of the published photometry has reached the main sequence.
Pal~14, at $\rgc \sim$70 kpc, was discovered by van den Bergh in 1958 and later 
classified as a GC by \citet{ar60}; it is also known as AvdB.
\citet{sa97} published the deepest CMD of Pal~14 to date, reaching more than a 
magnitude fainter than the main sequence turn off (MSTO), and concluded that Pal~14 is 
3--4 Gyr younger than inner halo GCs with similar metallicities.  

We describe the observations and data reduction in $\S$2 and present the WFPC2 CMDs
of AM-1 and Pal~14 in section $\S$3.  Available information regarding the reddening
and metallicity of both clusters is discussed in $\S$4 followed by relative and absolute
age analyses in $\S$5.  The results are discussed and put into context in $\S$6 and
the paper concludes with a summary in $\S$7.

\section{Observations and Data Reduction}

The observations used in the present study were obtained with HST/WFPC2 as part
of program number GO-6512 (PI: Hesser) during cycle 6. Table \ref{obslog} shows
the observing log. The 36 AM-1 and 14 Pal~14 images were retrieved from the HST archive 
and calibrated using the pipeline bias and flat-field procedures.

The stellar photometry was performed with the HSTPhot software \citep{dolphin} which
is designed especially for WFPC2 observations. A number of preprocessing steps were
performed before applying HSTPhot. First, image defects such as bad and saturated
pixels were masked using the data quality files available for each WFPC2 image. Cosmic
rays and hot pixels were flagged using tasks within HSTPhot. Next, the positional
offsets between each image and a reference image were measured using the IMEXAMINE
task in IRAF. These were input into the multiphotometry mode of HSTPhot in order
to measure the magnitudes of all detected profiles on each image. The measurements
were done using point spread functions (PSFs) that accompany the HSTPhot software
employing the `weighted PSF fitting mode' which places more weight on the
central pixels in a profile and less on the outer pixels. 

The instrumental magnitudes were transformed to the native WFPC2 VEGAmag system as 
well as the ground-based Johnson-Cousins VI system using equations included in HSTPhot. 
The final set of standardized photometry was extracted from the results file by 
selecting only those stars with 'Object Codes' that suggest a well-measured star-like 
profile. The object codes from HSTPhot are defined and described in the work of
\citet{dolphin}. These come from a comparison of the $\chi$ values of the fits using a 
stellar profile, a single pixel without a background, and a flat profile representing 
an extended object. For detections where the stellar profile fits best, one of three star 
classes is assigned to the object that characerize the probability that the object is 
either: a single star (object code = 1), an unresolved binary (object code = 2), or a
line-of-sight binary (object code = 3). In this work, we have selected only those 
detections with object code equal to one for further analysis.

\begin{figure}
\plotone{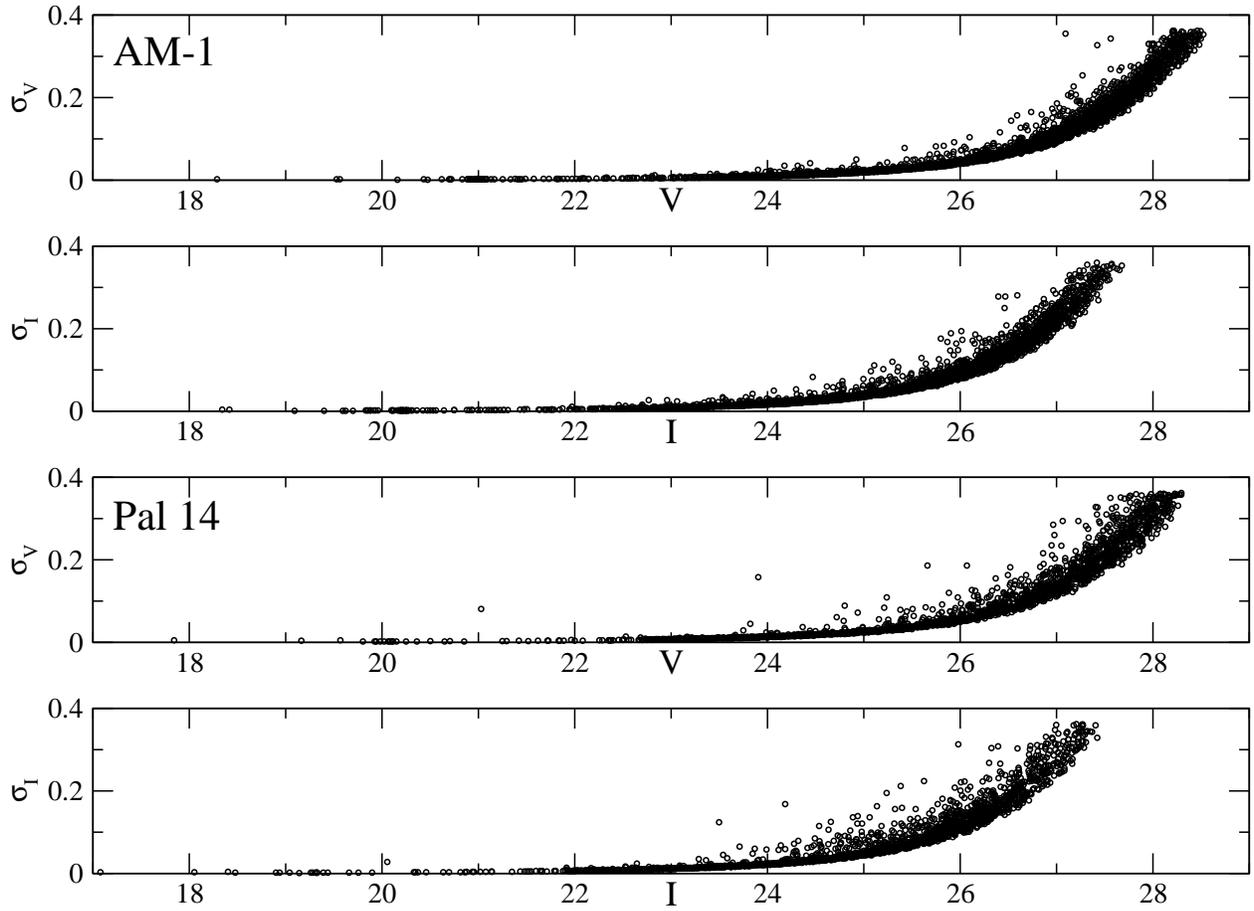}
\caption{Photometric errors as a function of apparent magnitude in AM-1 and Pal~14.
For the sake of brevity, we have used V=F555W and I=F814W in the plot.\label{photerr}}
\end{figure}

Figure \ref{photerr} shows how the photometric errors in both bands behave as a function
of apparent magnitude. In the plot, F555W and F814W are referred to as V and I, respectively.
For AM-1 the average photometric errors are 0.002 in both bands at the level of the 
HB and rise to 0.011 mag in F555W and 0.012 mag in F814W at the level of the MSTO.
For Pal~14 the average photometric errors are 0.002 mag in both bands at the level of the HB 
and rise to 0.007 mag in F555W and 0.008 mag in F814W at the level of the MSTO.  As Figure
\ref{photerr} demonstrates, the scatter in photometric errors is somewhat greater in Pal~14
than in AM-1.

\section{Color-Magnitude Diagrams}

\begin{figure}
\plotone{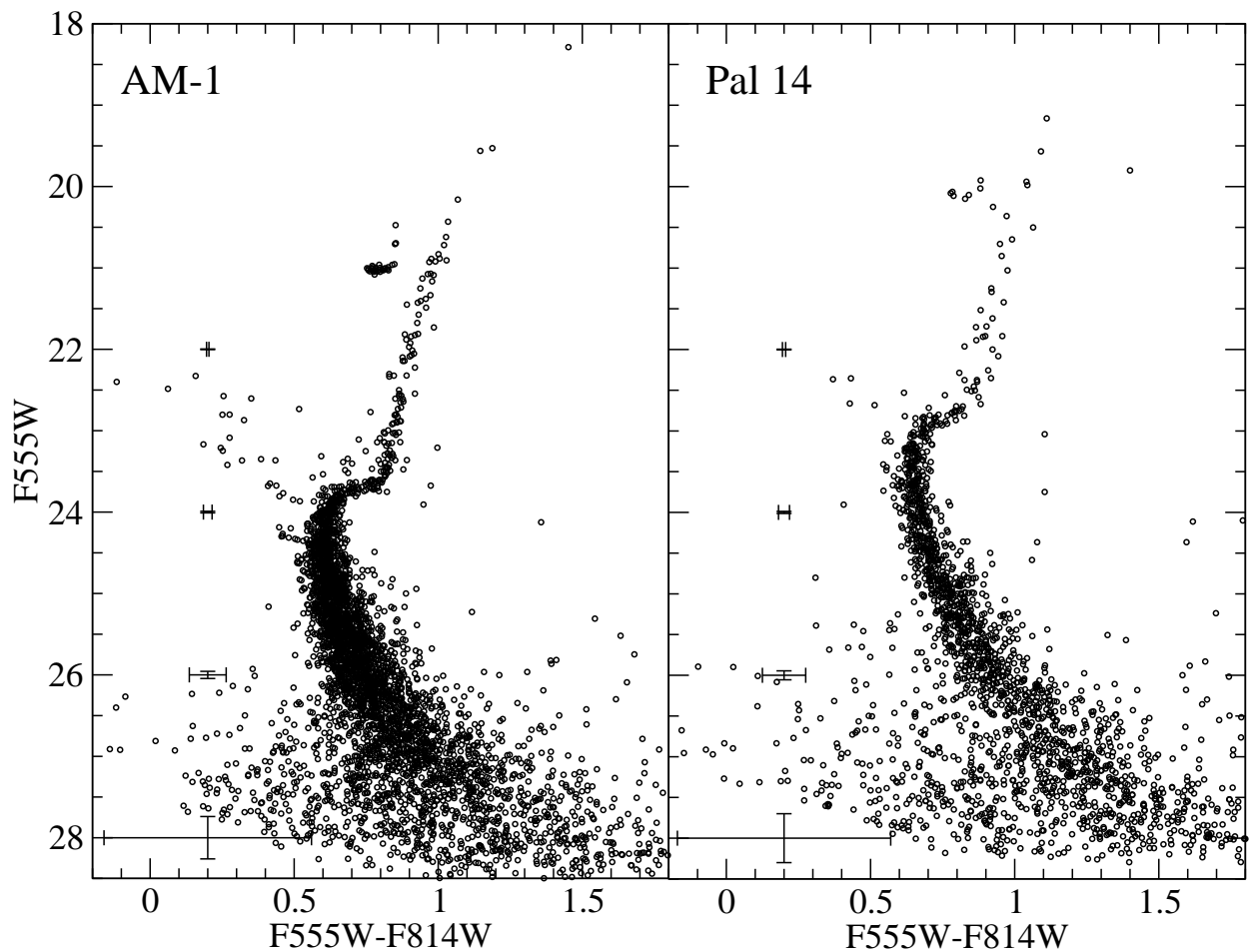}
\caption{HST/WFPC2 CMDs of AM-1 (left) and Pal~14 (right). AM-1 reveals a tight red 
clump, a well-defined RGB below the level of the red clump, a MS that extends $\sim$4 
mag below the turnoff, and an abundant collection of $\sim$20 blue stragglers.  Pal~14 
shows a sparsely populated red clump and RGB, a narrow MS that extends $\sim$5 mag 
below the turnoff, and a handful of blue stragglers. Characteristic photometric errors 
are plotted at intervals in both panels.\label{CMD}}
\end{figure}

The WFPC2 F555W--F814W CMDs of AM-1 and Pal~14 are presented in Figure \ref{CMD}.  
Both CMDs are remarkably free of field star contamination. Figure \ref{CMD} also indicates 
the average photometric errors by showing error bars at F555W=22, 24, 26, and 28 towards 
the left side of each panel.
The CMD of AM-1 (left panel of Figure \ref{CMD}) includes over 4,000 stars and extends 
from the brightest giant at F555W $\sim$ 18.5 down to approximately F555W = 29 though 
the main sequence becomes washed out by F555W $\sim$ 27.  The figure shows a 
well-populated MS and lower red giant branch (RGB), a tight red clump at F555W $\sim$ 21,
and a significant population of blue stragglers.  
The right panel of Figure \ref{CMD} shows over 2,000 stars in Pal~14. It exhibits a narrow MS 
that extends about 5 mags below the turnoff.  Stars brighter than the turnoff are 
relatively rare in Pal~14 and thus the red clump and RGB are not as well-defined as in AM-1.
Pal~14 contains a handful of blue stragglers.

\begin{figure}
\epsscale{0.9}
\plotone{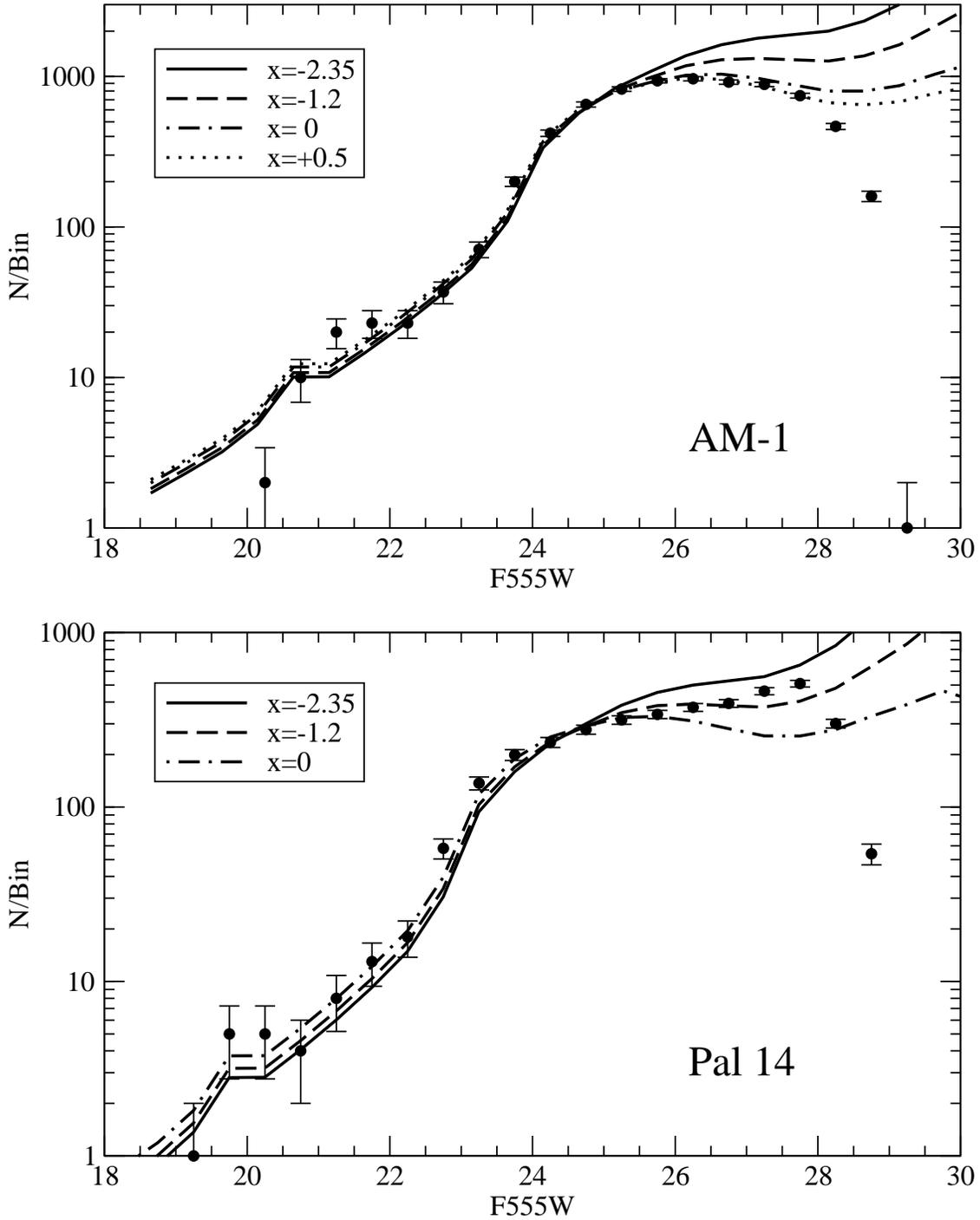}
\caption{Luminosity functions in F555W of AM-1 (top) and Pal~14 (bottom) with $\sqrt{N}$ error bars but without correcting for incompleteness.  Model LFs are plotted based on the isochrone fits described in $\S$5.2 and assuming power-law mass functions (which we define as dN/dM$\propto$M$^x$). AM-1 is fit well down to V=27 by a flat or positively-sloped MF but Pal~14 is rather poorly fit by the models, see text for discussion.\label{LF}}
\end{figure}

Figure \ref{LF} presents F555W luminosity functions (LFs) of AM-1 and Pal~14 but we stress 
that these results are preliminary and not corrected for incompleteness.  The error bars
are $\sqrt{N}$. Plotted along with 
the observed LFs are model LFs based on the isochrone fitting described in $\S$5.2 and 
assuming power-law mass functions (MFs).  In this paper, the power-law MF takes the form 
dN/dM$\propto$M$^x$ where x is referred to as the power-law slope.
The model LFs were normalized to the clusters by matching the number of stars within
$\pm$1 mag of the MSTO.  In the top panel, the AM-1 LF is fit well by models with a
flat (x=0) or positive power-law MF.  In the bottom panel, the LF of Pal 14 indicates a
negatively sloped MF but here the model fits are worse than for AM-1.  This is
related to the flat appearance of the Pal 14 LF (in the logarithmic plot) between
V=23 and ~28.  The models, regardless of power-law slope, suggest more curvature in this
region. The unusual LF of Pal 14 deserves further attention and will likely be improved 
by data covering more of the cluster.



\section{Reddening and Metallicity}

Before we can measure relative cluster ages we need to adopt a reddening and metallicity
for each cluster in our sample, including our comparison cluster, for which we
have selected the well-known Galactic globular M~3.  
No single spectroscopic study encompasses AM-1, Pal~14, and M~3. However, with M~3 measured
by \citet[hereafter ZW84]{zw84}, AM-1 measured by \citet{su85} using the same spectral 
features as ZW84, and Pal~14 measured by \citet{ar92}, there is something 
approacing a common scale. \citet{ar92} found good agreement between their [Fe/H] values and 
ZW84 for clusters that appeared in both studies.

Table \ref{param} shows the distance moduli from \citet{ha96}, reddening estimates from the
\citet{sc98} dust maps, metallicities from the sources cited in the previous paragraph, and 
the iron and $\alpha$-element abundances for M~3 from \citet[hereafter PVI]{pvi}.
PVI determined abundances by averaging high resolution spectroscopic measurements from 
36 stars in M~3. The PVI abundances are likely the most accurate available for M~3
and were adopted for the purpose of isochrone fitting in $\S$5.
Since the metallicities listed in column three of Table \ref{param} for each cluster agree 
to within 0.1 dex, and certainly to within the quoted errors, it is reasonable to 
differentially compare these three GCs.
It should also be noted that \citet{hi06} determined that a Yonsei--Yale isochrone \citep{ki02} 
with [Fe/H] = --1.4 and [$\alpha$/Fe] = +0.3 at 11 Gyr provided the best fit to the RGB.
Since the photometry presented by \citet{hi06} for AM-1 did not reach the MSTO 
we do not consider this result---at least regarding the age---to be well--constrained.


Since the observational data considered in this paper are in either the ground-based V--I or
WFPC2 F555W--F814W color and the reddening estimates are given in E(B--V) it is necessary to 
convert them. \citet{bcp98} determined E(V--I)/E(B--V) = 1.32 + 0.06(V--I) for stars with 
spectral types from B to K.  Since we are interested in colors lying between 0.5 $<$ V--I 
$<$ 1.5, we chose (V--I) = 1 as the average color and thus adopted E(V--I) = 1.38~E(B--V).  
In the native HST/WFPC2 photometric system, we take E(F555W--F814W) = 1.2~E(B--V) from
Table 13 of \citet{ho95}.

\section{Cluster Age Determinations}

\subsection{The Horizontal Method}

We begin by employing the relative age-dating technique known as the
horizontal method pioneered by \citet{sd90} and \citet[hereafter VBS]{vbs90}. 
Adopting the approach advocated by VBS (also used by S99), we register the 
photometry of each program
cluster and the fiducial sequence of a comparison cluster with similar metal
abundance using the color of the MSTO and the magnitude
at a point on the main sequence that is 0.05 mag redder than the MSTO. Once
registered in this manner, younger clusters will have redder RGBs than older 
ones. 

\begin{figure}
\plotone{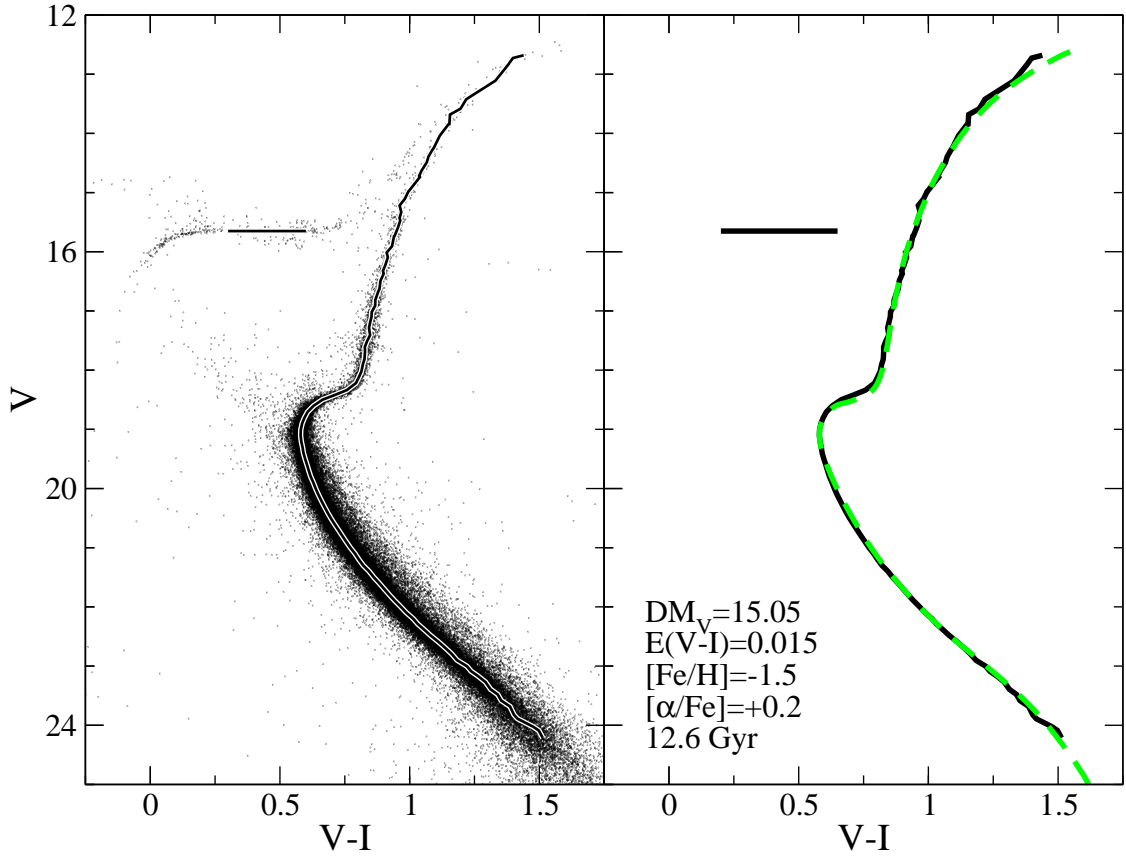}
\caption{The left panel shows the HST/ACS CMD of M~3 with the fiducial sequence (solid line) over-plotted. The CMD only includes stars with photometric errors less than 0.05 mag in both bands.  The right panel compares the best fit isochrone (dashed line) to the fiducial sequence (solid line).\label{M3fid}}
\end{figure}

For our comparison cluster, M~3, 
the CMD used to construct a fiducial sequence was taken from the ACS Survey of 
Galactic GCs \citep{sa07}.  For direct comparison with the data presented in this paper, 
the ACS data were converted from F606W and F814W to ground-based V and I using 
the transformation equations in \citet{si05}.  A fiducial was constructed using only stars
with photometric errors less than 0.05 mag in both bandpasses.  The high quality CMD 
and the fiducial sequence of M~3 are shown in the left panel of Figure \ref{M3fid}. 
We now proceed to compare this sequence to the photometric data of our
program clusters.

\begin{figure}
\plotone{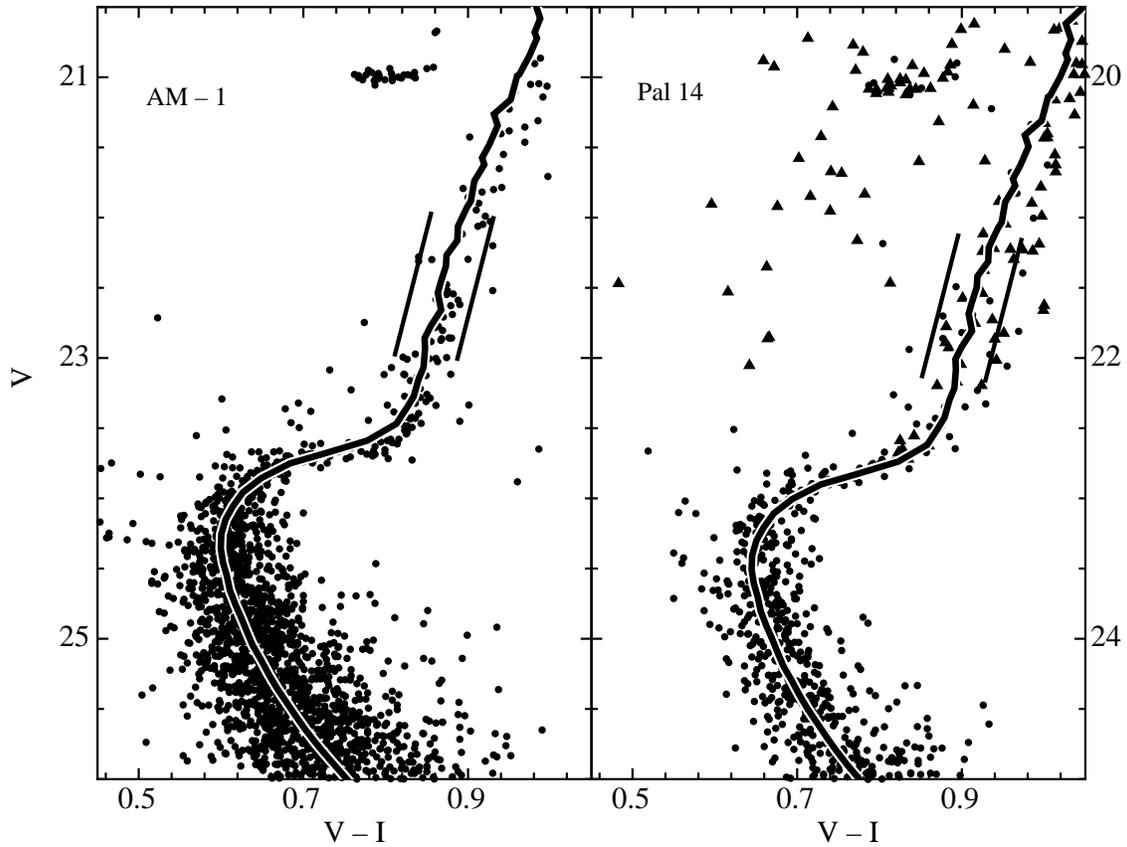}
\caption{The horizontal (VBS) method applied to our photometry of AM-1 (left panel) and
Pal~14 (right panel). The fiducial sequence (solid line) is that of the globular cluster M~3. 
The filled triangles in the right panel represent the ground-based
Pal~14 photometry obtained from \citet{st00} to which the WFPC2 photometric zero point has been offset. The parallel lines along the lower RGB represent age differences of $\pm$2 Gyr.\label{vbs}}
\end{figure}

The left panel of Figure \ref{vbs} shows our photometry for AM-1 and the 
fiducial sequence of M~3 matched using the prescription of VBS while
the right panel of the same figure illustrates the comparison of Pal~14 with
M~3 matched in the same manner. Note that the evolved stars in Pal~14
are relatively few in number (see Figure \ref{CMD}). As a result, we have 
augmented the WFPC2 photometry (circles) with ground-based photometric standard 
stars reported by \citet[triangles]{st00}. We have offset the WFPC2 data in color 
and magnitude to match the photometric zero point of the ground-based data. 
Specifically, 

\begin{eqnarray}
<V(HST) - V(Stetson)> = 0.068 \pm 0.007;  N = 31\\
<I(HST) - I(Stetson)> = 0.007 \pm 0.007;  N = 17
\end{eqnarray}

The lines parallel to the fiducial RGBs represent the locations of clusters that are 
2 Gyr older (left) and 2 Gyr younger (right) than the comparison cluster. 
The $\pm$2 Gyr lines were derived by applying the VBS method to [Fe/H]=--1.5,
[$\alpha$/Fe]=+0.2 isochrones from \citet{d08} with ages ranging from 10 to 14 Gyr.
From the isochrones we found that V--I decreases by 0.0197 mag per Gyr and thus
the lines were plotted at $\sim$0.04 mag on either side of the M~3 fiducial.
 
The RGBs of AM-1 and Pal~14 suggest that these clusters are 1--2 Gyr younger
than M~3. Figure \ref{vbs} shows that the RGB stars in AM-1 lie to the red of the M~3
fiduical sequence.  The case of Pal~14 is less concrete due to the lack of a well-defined
sequence of stars. The trend in Pal~14 is more obvious if
one considers the range from V=20 to 22. These relative age statements depend 
on the assumption that the program clusters and M~3 share similar chemical compositions.

\subsection{Isochrone Fitting}
Before the isochrone fitting results are presented, two points regarding the CMDs 
of AM-1 and Pal~14 need to be made. First, we have elected to present the isochrone
analysis of the AM-1 CMD in the native HST/WFPC2 system
because no published ground-based VI photometry exists for this cluster, which
would allow us to check our photometric zero point.  
Second, as we did in the application of the horizontal method in the previous section,
we will use the ground-based Pal~14 data to better define the location of the
evolved-star sequences.

Theoretical isochrones from the Dartmouth Stellar Evolution Database
\footnote{http://stellar.dartmouth.edu/$\sim$models/}\citep{d07,d08} 
were used in the analysis. These isochrones employ state of the art input physics and
include partially inhibited diffusion and gravitational settling of He and metals 
\citep{c01} and a model atmosphere based surface boundary condition. 
The reader is referred to \citet{d07,d08} for a full discussion of their properties.
These models have been tabulated in 0.5 Gyr intervals but, for the purpose of
fitting isochrones to fiducials and CMDs, the isochrones have been interpolated to 
produce a finer grid spacing of 0.1 Gyr in order to allow for more 
precise age constraints.

\begin{figure}
\plotone{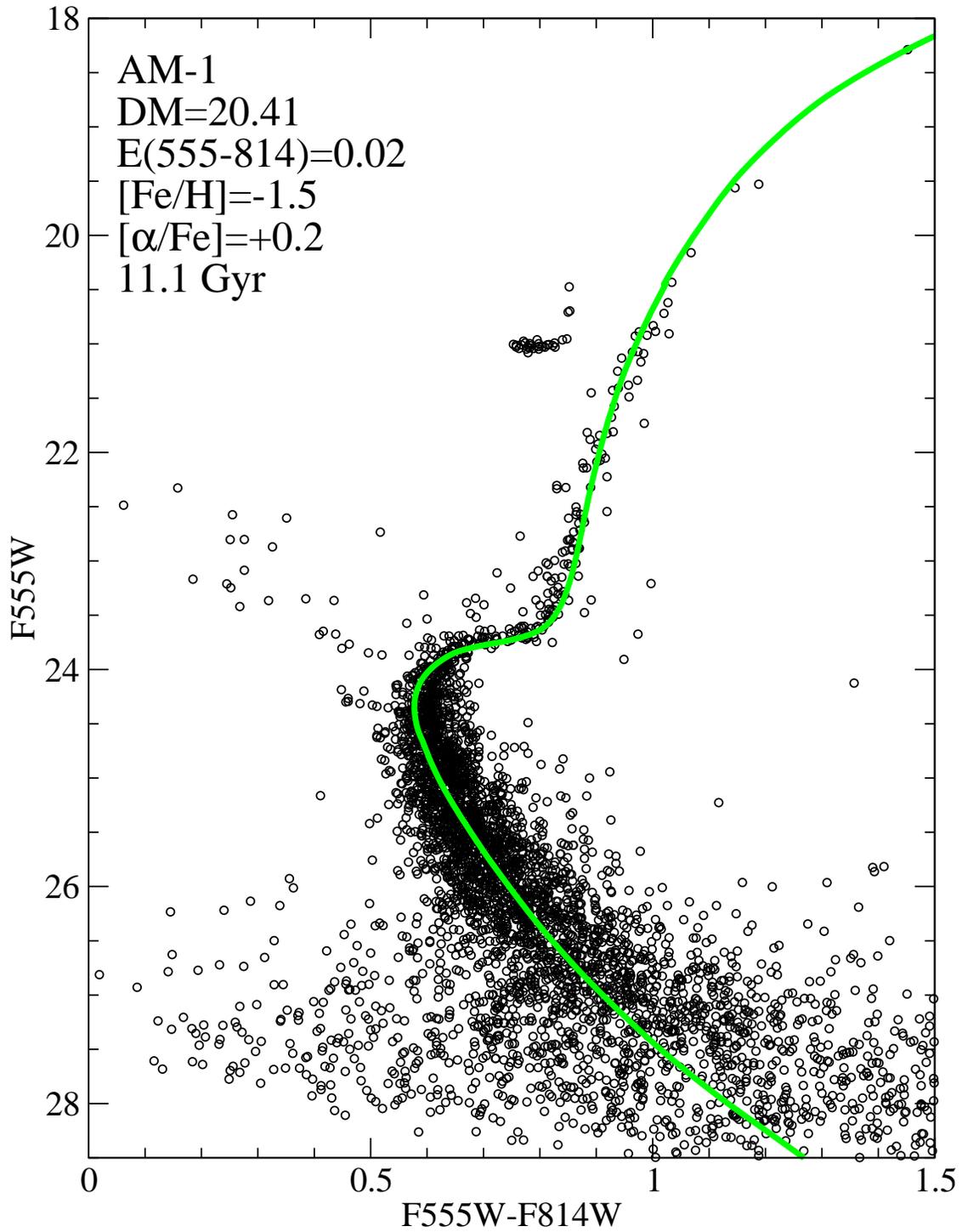}
\caption{Results of isochrone fitting to AM-1.  The relevant parameters used in the fit are listed on the figure.\label{AM1iso}}
\end{figure}

\begin{figure}
\plotone{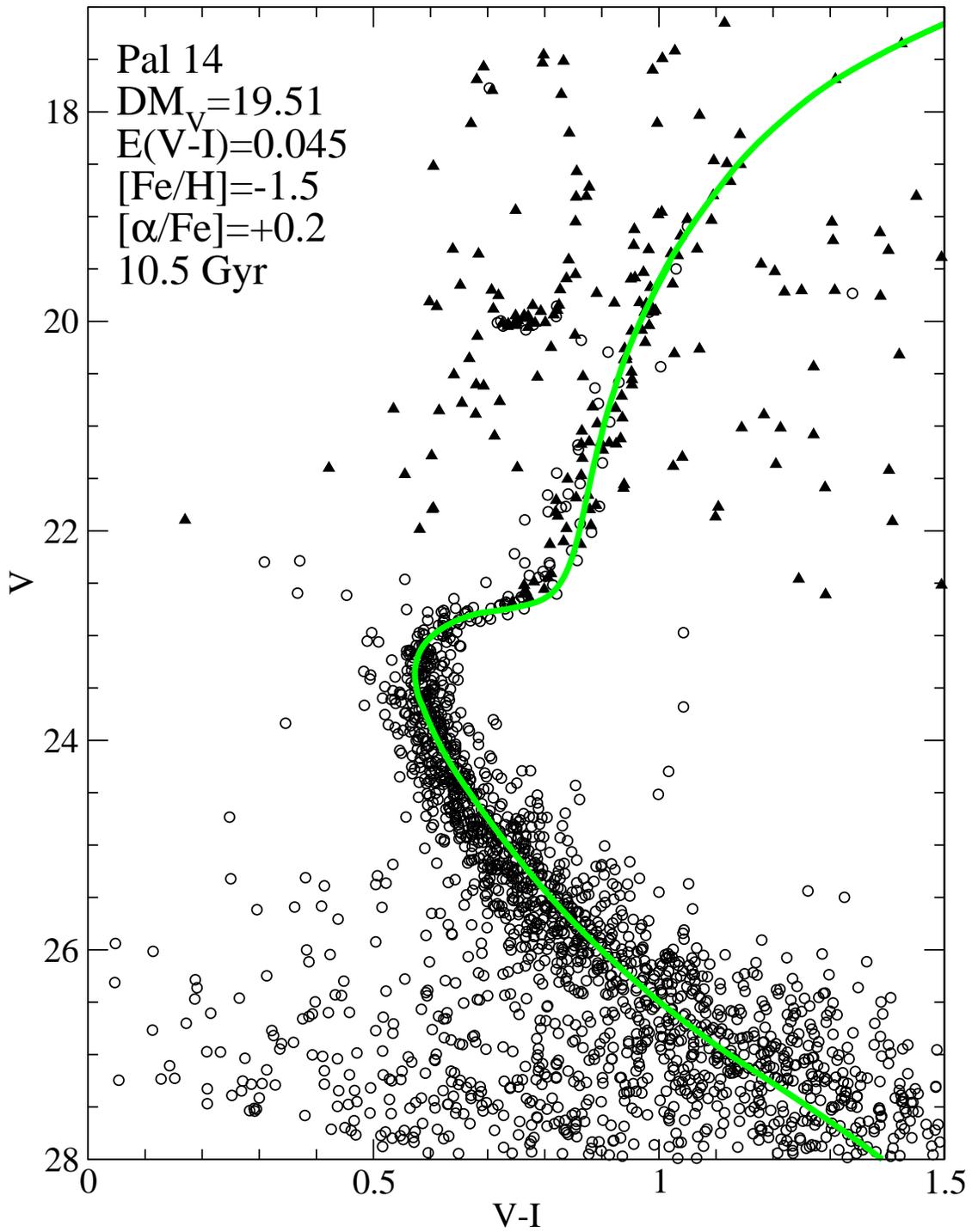}
\caption{Same as Figure \ref{AM1iso} but for Pal~14. The open circles are WFPC2 photometry and the filled triangles are from \citet{st00}.\label{Pal14iso}}
\end{figure}

Adopting the PVI abundances of M~3 and starting with the distance and reddening estimates 
in Table \ref{param}, we then vary the age, distance, and reddening until a satisfactory
correspondence is achieved between the main sequence of the cluster and the isochrone, 
making sure that the point 2 mag below the MSTO (i.e., $M_V \sim$ 6) matches the data. 
This is the same procedure followed by \citet{sa07}. 
The results of isochrone fitting are collected in Table \ref{ages} and shown in 
the right panel of Figure \ref{M3fid}, Figure \ref{AM1iso}, and Figure \ref{Pal14iso}.
In short, we find that Pal~14 is younger than AM-1 by $\sim$0.5 Gyr if both clusters
have similar abundances.

The distance and reddening estimates
derived from the isochrone fits provides a check on the fitting method.  Comparing the 
reddening estimates derived from isochrone fitting in Table \ref{ages} to the values 
listed in Table \ref{param}, we find that differences in the reddening values are
within the expected uncertainty in a given reddening of $\pm$0.02 mag in E(B--V). 
As with the reddenings, the isochrone-based distances listed in Table \ref{ages}
generally agree with those in Table \ref{param} to within the expected error
of $\pm$0.1 mag in distance modulus.

\subsection{Consistency Check on the Ages}

Thus far we have presented two types of age analyses, the horizontal VBS method and
isochrone fitting, to probe the ages of AM-1 and Pal~14 relative to M~3. Both
methods suggest that the program clusters, which exhibit HBs that are 
predominantly redward of the RR Lyrae instability strip, are 1--2 Gyr younger 
than the comparison cluster, M~3, which exhibits stars on the red and
blue portions of the HB. In this section, our aim is to provide a consistency
check of those results using the method employed by \citet{sll}.  This method
compares the relative locations of the MSTO in two clusters whose CMDs have 
been properly registered as described in more detail below.

Since all three clusters have similar metallicities, we expect them to have 
comparable HB absolute magnitudes and RGB colors at the level of the HB.
This is slightly modulated by the fact that cluster age also affects the level
of the red clump and the color of the RGB. The former has been calibrated by
\citet{cr01} to be 0.028 mag/Gyr for ages older than $\sim$5 Gyr and 
metallicities below [Fe/H] $\sim$ --1.0. The latter can be assessed using our
isochrones from which we find a variation of 0.003 mag/Gyr in the V--I 
color of the RGB at the level of the HB. 

\begin{figure}
\epsscale{1.0}
\plotone{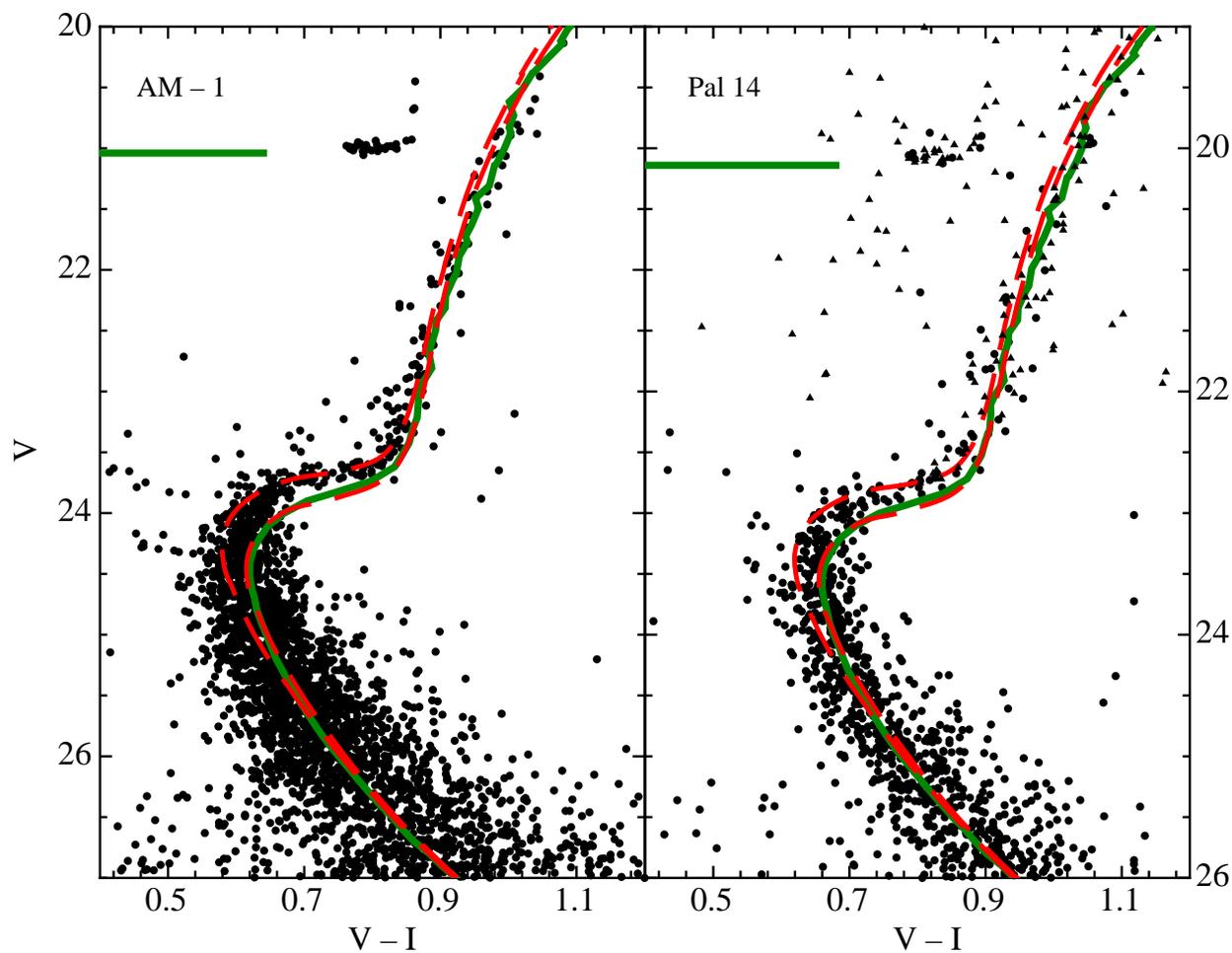}
\caption{The \citet{sll} method applied to AM-1 (left panel) and Pal~14 (right panel). 
The M~3 fiducial sequence (solid lines) and isochrones with [Fe/H]=--1.5, [$\alpha$/Fe]=+0.2, and ages of 10.6 and 12.6 Gyr (dashed lines) are plotted. The filled triangles in the right panel represent the ground-based Pal~14 photometry obtained from \citet{st00} to which the WFPC2 photometry (circles) has been matched.\label{mstocomp}}
\end{figure}

With these guidelines in mind, we proceed to use the measured HB
magnitudes and RGB colors of AM-1, Pal~14, and M~3 along with their
relatives ages (Table \ref{ages}) to shift them to the same distance and reddening
scale. This allows us to compare their MSTO/SGB locations
with each other as shown in Fig. \ref{mstocomp}. We see that the turnoff
regions of both AM-1 and Pal~14 are demonstrably brighter than M~3.
Also included  in Fig. \ref{mstocomp} are \citet{d08} isochrones with [Fe/H]=--1.5
and [$\alpha$/Fe]=+0.2 at 10.6 and 12.6 Gyr (the best-fit age to the M~3 fiducial and
2 Gyr younger). The MSTO regions 
of AM-1 and Pal~14 are consistent with ages that are 1--2 Gyr younger than M~3, in 
agreement with the horizontal method and isochrone fitting.

As we close out this section, it is important to reiterate that the age estimates 
presented in this paper are accurate only to the extent that AM-1, Pal~14, and M~3 
have similar compositions.  In order to gauge the effect of changing either [Fe/H] or 
[$\alpha$/Fe] we use an isochrone with [Fe/H]=--1.5 and [$\alpha$/Fe]=+0.2 at 12 Gyr 
as the baseline and search for an isochrone with the same separation between the MSTO 
and the base of the RGB at different compositions. An increase (decrease) in [$\alpha$/Fe] 
of 0.2 dex causes the age to decrease (increase) by 0.5 Gyr.  If [Fe/H] increases by 0.2 
dex, then the age decreases by 0.5 Gyr but if [Fe/H] decreases by the same amount, then 
the age increases by $\sim$1 Gyr.  Our age measurements must be viewed with some caution 
until more conclusive abundances can be derived for the sample clusters.

\section{Discussion}

We provide strong evidence that AM-1 and Pal~14 join the S99 sample of Pal~3, Pal~4, and
Eridanus in having younger ages than comparable inner halo GCs.  Hence all five GCs
with $\rgc >$ 50 kpc and red HBs are younger than their counterparts with bluer HBs in
the inner halo.
These five GCs are second parameter clusters (in the sense that they have redder HBs
at a given metallicity than the inner halo trend would suggest) and all have been
demonstrated to have relatively young ages. This makes a strong case for age as the
second parameter in the outer halo. Can we make a case for age as the second parameter
throughout the Galaxy - i.e. among the entire sample of Galactic globular clusters?

\citet[see their Table 1]{m05} presented a compilation of parameters (e.g. $\rgc$,
HB type, [Fe/H]) for the entire Galactic GC population. In addition, each is assigned
to a sub-population based on the categories introduced by \citet{z93} along with the
addition of two categories not available in 1993: GCs associated with the Sagittarius
dwarf galaxy and GCs containing multiple stellar populations. According to the
\citet{z93} classification, GCs that follow the inner halo trend of HB type vs. [Fe/H]
belong to the ``Old Halo'' (OH) while those that diverge from this trend by a certain
amount (or more) belong to the ``Younger Halo'' (YH).

The YH clusters are synonymous with the second parameter phenomenon and while
they dominate the OH population at large $\rgc$, the vast majority of them have
$\rgc <$ 50 kpc. Thus the outer halo clusters may not be a representative sample of 
the entire YH population but there is evidence for age as the second parameter among GCs 
with $\rgc <$ 50 kpc as well. Among these, GCs such as Ruprecht~106 \citep{buon93}, IC~4499
\citep{fer95}, Palomar~12 \citep{stet89}, and Pyxis \citep{sarageis96} have been shown to be 
younger than inner halo GCs with similar metallicities and bluer HB morphologies. This trend 
is also seen through the detailed examination of `second parameter pairs' such as NGC~288 and
NGC~362 \citep[VBS]{sd90} or M~3 and M~13 \citep{rey2001}.  At the same time, accurate
abundance determinations of second parameter GCs (particularly the more distant
ones) are desperately needed. Accurate abundances for the YH population, coupled with
a systematic age analysis of these clusters,  will provide deeper insight into the
Galactic GC population and the nature of the second parameter. 

We note in passing that \citet{m05} do not consider Pal~12 to be a member of the YH population 
because of its supposed association with the Sgr dwarf spheroidal galaxy. However, Pal~12 is 
spatially distinct from the main body of Sgr so that we can consider it to be a member of the 
Milky Way halo. In any case, if the hypothesis of \citet{sz78} and the implications of the 
analysis performed by \citet{m05} are correct---that all of the YH clusters originated in 
(now disrupted) dwarf spheroidal galaxies---then we have no choice but to consider clusters 
like Pal~12 part of the Milky Way's YH population and thus part of the Milky Way halo.

\section{Conclusion}
Color magnitude diagrams of AM-1 and Pal~14 from HST/WFPC2 F555W--F814W photometry were presented.
The CMDs reveal unprecedented depth, reaching $\sim$4 mag below the MSTO in both clusters.  
AM-1 and Pal~14 have very similar CMD morphologies.  AM-1 shows a more densely populated RGB and HB, 
albeit with very few bright red giants, and an impressive display of blue stragglers.  Pal~14 has a 
sparsely populated RGB and HB which is likely due to its large size (compared to the area covered by
WFPC2) and low density.

Three different age measurement techniques reveal that AM-1 and Pal~14 are $\sim$1.5--2 
Gyr younger than M~3 assuming all three GCs have similar compositions. Isochrone fitting
suggests that if AM-1 and Pal~14 have similar abundances, then Pal~14 is about 0.5 Gyr
younger than AM-1.  S99 found that
Pal~3, Pal~4 and Eridanus are also $\sim$1.5--2 Gyr younger than inner halo GCs of 
comparable metal abundance.  Thus all five of the GCs with $\rgc >$ 50 kpc and red HBs 
are 1.5--2 Gyr younger than GCs with similar metallicities in the inner halo (to the
extent that assumptions regarding similar abundances are valid).

The variation of the second parameter phenomenon with $\rgc$ has been known for some time, but 
the identity of the second parameter itself remains a topic of active debate in the literature, 
even some 40 years after it was discovered.  When combined with previously published results on 
the ages of clusters that exhibit the second parameter effect, the present study provides further 
evidence that age is a strong candidate for the second parameter. However, there are a significant 
number of globular clusters important to identifying the second parameter that have not been 
accurately age-dated.  Of equal importance is the precise determination of abundances in these 
clusters.  Improvements to both the total metal abundance and the abundance ratios in these clusters
will lead to more appropriate comparisons with well-studied, nearby GCs and theoretical 
models and, ultimately, to a better understanding of the second parameter phenomenon and
the formation history of the Galactic halo.

\acknowledgments
We thank Peter Stetson for sharing the HST photometry of Pal~3, Pal~4, and Eridanus.
We thank the anonymous referee for insightful comments that improved the presentation and 
focus of the paper.

\begin{deluxetable}{lccc}
\tablecolumns{5}
\tablewidth{0pc}
\tablecaption{Observing Log\label{obslog}}
\tablehead{\colhead{Cluster}&\colhead{Date}&\colhead{Filter}&\colhead{Exposure Time}
}
\startdata
AM--1     & 1999 June & F555W & 4 x 180s     \\
                &                     &               & 1 x 900s     \\  
                &                     &               & 8 x 1000s     \\
                &                    &               & 5 x 1300s     \\
                &                    & F814W & 4 x 230s    \\
                &                    &                & 5 x 1200s \\
                &                    &               &  4 x 1300s \\
                &                    &                & 5 x 1400s \\
Palomar 14     &  1999 July  & F555W & 4 x 160s     \\
                          &                     &               & 1 x 900s     \\  
                          &                     &              & 7 x 1000s     \\
                          &                    & F814W & 4 x 230s    \\
                          &                    &                & 4 x 1100s \\
                          &                    &               &  2 x 1200s \\
                          &                    &                & 2 x 1300s \\
\enddata
\end{deluxetable}

\begin{deluxetable}{cccccc}
\tablecolumns{6}
\tablewidth{0pc}
\tablecaption{Globular Cluster Parameters\label{param}}
\tablehead{\colhead{Name}&\colhead{$DM_V$}&\colhead{E(B--V)}&\colhead{[Fe/H]}&
\colhead{[Fe/H]$_{PVI}$}&\colhead{[$\alpha$/Fe]$_{PVI}$}}
\startdata
M~3     & 15.12 & 0.013 & --1.66$\pm$0.06 & --1.50 & +0.22   \\
AM-1    & 20.43 & 0.008 & --1.7$\pm$0.2  & \nodata& \nodata \\
Pal~14  & 19.47 & 0.035 & --1.60$\pm$0.18 & \nodata& \nodata \\
\enddata
\end{deluxetable}

\begin{deluxetable}{ccccc}
\tablecolumns{5}
\tablewidth{0pc}
\tablecaption{Isochrone-based Parameters\label{ages}}
\tablehead{\colhead{Name}&\colhead{DM$_V$}&\colhead{E(B--V)}&\colhead{Age (Gyr)}&\colhead{Rel. Age (Gyr)}}
\startdata
M~3     & 15.05 & 0.011  & 12.6 & \nodata \\
AM-1    & 20.41 & 0.016 & 11.1 & --1.5   \\
Pal~14  & 19.51 & 0.033 & 10.5 & --2.1   \\
\enddata
\end{deluxetable}

\end{document}